\documentclass[preprint,aps]{revtex4}
\usepackage{epsfig}

\newcommand{\beq}{\begin{equation}}
\newcommand{\eeq}{\end{equation}}
\newcommand{\be}{\begin{eqnarray}}
\newcommand{\ee}{\end{eqnarray}}

\begin{document} 


\title{
CP violation in the associated production of a charged Higgs boson
with a top quark at the CERN Large Hadron Collider
}

\author{Dong-Won Jung,$~~$ Kang Young Lee
\thanks{kylee@muon.kaist.ac.kr}}

\affiliation{
Department of Physics, 
Korea Advanced Institute of Science and Technology, \\
Daejeon 305-701, Korea}

\author{H.S. Song}

\affiliation{
        School of Physics,
        Seoul National University, Seoul 151-742, Korea
}

\date{\today}
\vspace{2cm}

\begin{abstract}
We explore the CP violation in the charged Higgs production
associated with a top quark at the LHC in the MSSM.
The supersymmetric phases of gaugino masses and trilinear $A$ terms
lead to a CP violation of the cross section
in the $ p p \to g b \to t H^\pm$ process
through loop corrections to the $tbH^\pm$ vertex.
We find that more than 10 \% of CP violating asymmetry is possible 
when the charged Higgs boson is heavy enough.
\end{abstract}

\vspace{2cm}

\maketitle






The Higgs mechanism provoked by the non-zero vacuum expectation value (VEV)
of a scalar doublet is the key structure of the Standard Model (SM),
which gives rise to the electroweak symmetry breaking
and the generation of fermion masses.
There is one SU(2) doublet of scalar field in the SM lagrangian
and only one neutral Higgs boson exists as a physical state
after the breakdown of the electroweak symmetry.
The Higgs boson is the only unobserved ingredient of the SM.
LEP experiment favors a light Higgs boson from the global fit of data 
and has suggested a candidate of the Higgs boson of 115 GeV \cite{higgs1}.
The discovery of the Higgs boson is the principal goal
of the CERN Large Hadron Collider (LHC).
For the SM Higgs boson, LHC promises to cover the wide range of
Higgs boson mass \cite{higgs2}.

Most extensions of the SM require the introduction of 
extended Higgs sector to the theory.
Generically, charged Higgs boson arises in the extended Higgs sector,
which does not exist in the SM.
It implies that the observation of a charged Higgs boson is
a clear evidence for existence of the new physics beyond the SM.
The minimal supersymmetric standard model (MSSM) is  
one of the most promising  candidate for physics beyond the SM.
In the MSSM, we need two Higgs doublets to preserve the supersymmetry 
and then there are a pair of charged Higgs boson $H^\pm$ with
two neutral CP-even Higgs bosons $h^0$, $H^0$,
and a neutral CP-odd Higgs boson $A^0$
in the particle spectrum of the MSSM.
Direct experimental search for $H^\pm$ at LEP II
has set the lower bound of mass to be 79.9 GeV \cite{lep2}
through $e^- e^+ \to H^- H^+$ production.
The direct and indirect studies of $H^\pm$ at the Tevatron
constrain the parameter space of $(m_{H^\pm}, \tan \beta)$.
CDF and D0 groups have excluded the low and high $\tan \beta$ regions 
up to $\sim 160$ GeV \cite{tevatron}.
The discovery potential of the charged Higgs boson at the LHC
has been studied by ATLAS \cite{atlas} and CMS \cite{cms} groups.
We expect to discover a charged Higgs boson as heavy as 1 TeV 
at the 5-$\sigma$ confidence level,
or may exclude it up to the mass of 1.5 TeV at 95 \% C.L. at the LHC
with the MSSM radiative corrections \cite{belyaev}.
We concentrate on the process $gb \to tH^\pm$ 
which is the most promising channels for the charged Higgs boson 
production at the LHC when $H^\pm$ is heavier than the top quark.
The discovery potential of this channel 
has been studied in the Ref. \cite{gbth1,gbth2}.
The Drell-Yan mechanism $gg$, $q \bar{q} \to H^- H^+$ 
and the associated production with a $W$ boson, 
$q \bar{q} \to H^\pm W^\mp$ are suppressed
due to the weak couplings, low quark luminosity, and
loop suppressions.
If the mass of $H^\pm$ is below the top quark mass,
$H^\pm$ is produced by the sequential decay of top quark 
$t \to b H^\pm$ following the $t \bar{t}$ pair production.

Exotic CP violating phenomena are generically predicted
in the MSSM due to new complex phases. 
Assuming the R-symmetry and the gauge coupling unification at the GUT scale,
the physical phases come from the $\mu$ parameter, 
the soft trilinear coupling $A$ terms, and two gaugino masses
after rotating away other phases. 
Then the supersymmetric loop corrections to the $tbH^\pm$ vertex 
involving squarks, charginos, gluinos and neutralinos 
generically lead to the manifest CP violation
in charged Higgs production and decay
due to the SUSY CP phases.
The CP violation in the decays of $H^\pm$ into $t,b$ quarks
and $\tau,\nu$ has been studied in Ref. \cite{christova,christovatau}.
In this work, we investigate the direct CP violating asymmetry
in the MSSM charged Higgs boson production 
in association with a single top quark at the LHC.



In the MSSM, the relevant interaction lagrangian of 
the charged Higgs boson production is given by
\be
{\cal L}_{\rm int} = V_{tb} \bar{t} \big( Y_t^+ P_L + Y_b^+ P_R \big) b H^+
                     + H.c.,
\ee
ignoring the light quark masses and mixing.
We incorporate the SUSY correction in this Yukawa coupling
to make the effective vertex.
The SUSY loop corrections arise through
the vertex corrections with
$\tilde{t}-\tilde{b}-\tilde{g}(\chi^0)$ loop,
$\chi^\pm-\chi^0-\tilde{t}(\tilde{b})$ loop,
$H^0 - W^\pm - t(b)$ loop,
and the $H^\pm - W^\pm$ self-energy diagram with
$\chi^\pm-\chi^0$ loop, $\tilde{t}-\tilde{b}$ loop, 
$H^0 - W^\pm$ loop.
We parametrize the Yukawa couplings as
$ Y_{t,b}^\pm = y_{t,b}^0 + \delta y_{t,b} 
            \pm \frac{1}{2} \delta y_{t,b}^{\rm CP}, $
where 
the SUSY loop corrections consist of the CP invariant part
$\delta y_{t,b}$ and the CP violating part $\delta y_{t,b}^{\rm CP}$.
The relevant quantity for the CP violation 
is ${\rm Re}~\delta y_i^{\rm CP} \equiv {\rm Re}~(Y^+_i-Y^-_i)$
where $i=t,b$ of which analytic form is derived in Ref. \cite{christova}
for the decay of the charged Higgs boson.
Since we consider the one loop correction to the vertex,
the effective vertex is expressed by one loop integral and,
in consequence by the Passarino-Veltman function.
We note that the $t$ ($\bar{t}$) quark of the $tbH^\pm$ vertices
is not on shell for the $t$-channel process 
and the argument of the Passarino-Veltman function
should be the partonic Mandelstam variable $\hat{t}$ instead of $m_t^2$.,
The principal contribution to ${\rm Re}~\delta y_i^{\rm CP}$ comes from 
$\tilde{t}-\tilde{b}-\tilde{g}(\chi^0)$ loop
and $\tilde{t}-\tilde{b}$ loop diagrams depicted in Fig. 1.
With the minimal supergravity model (mSUGRA) type particle spectrum,
contributions of $\chi^\pm-\chi^0-\tilde{t}(\tilde{b})$,
$H^0 - W^\pm - t(b)$, $\chi^\pm-\chi^0$ and
$H^0 - W^\pm$ loops to ${\rm Re}~\delta y_i^{\rm CP}$ 
are less than 1 \% for any value of the charged Higgs mass 
and irrelevant for observation in the experiment.


In terms of the couplings given in Eq. (1),
we write the scattering amplitude as
\be
i{\cal M}(g \bar b \to \bar t H^+)
= i g_s \bar{b} \left[ T_{ab}^c \not{\epsilon}(p_g)
          \frac{Y^-_t P_L + Y^-_b P_R}{\not{p}_H - \not{p}_b - m_t} 
        + \frac{Y^-_t P_L + Y^-_b P_R}{\not{p}_H + \not{p}_t - m_b} 
            \not{\epsilon}(p_g) T_{ab}^c
        \right]t,
\ee
and obtain the partonic cross section as
\be
\hat \sigma(g \bar{b} \to \bar{t} H^+) &=& 
\frac{1}{32 \pi \hat s}
\frac{\sqrt{(\hat s - m_t^2 - m_H^2)^2 -4 m_H^2 m_t^2}}{\hat s - m_b^2}
|\bar {\cal M} |^2,
\ee
where the colliding energy is kinematically allowed 
if $\hat s > (m_t + m_H)^2$. 
We define the direct CP asymmetry at the parton level 
\be
\hat A_{CP} 
= \frac{\hat \sigma( g \bar{b} \to \bar{t} H^+)
              -\hat \sigma( g b \to t H^-)}
              {\hat \sigma( g \bar{b} \to \bar{t} H^+)
              +\hat \sigma( g b \to t H^-)}
\approx 
\frac{y_t^0 {\rm Re}~\delta y_t^{\rm CP} 
    + y_b^0 {\rm Re}~\delta y_b^{\rm CP} }
     {|y_t^0|^2 + |y_b^0|^2 },
\ee
when assuming $\delta y_i^{\rm CP} \ll y_i^0$.
The relevant quantity for the CP asymmetry
${\rm Re}~\delta y_i^{\rm CP} \equiv {\rm Re} (Y^+_i-Y^-_i)$
consists of the supersymmetric CP violating phase part 
and imaginary part of the loop integrals such that
${\rm Re}~\delta y_i^{\rm CP} \propto$ Im(squark couplings)$\cdot$
Im($C_I,~B_0$), $I=0,1,2$ where $C_I,~B_0$ are 3-point and 2-point
Passarino-Veltman functions \cite{passarino}.
The SUSY phases in trilinear $A$ terms, $\mu$ parameter
and gaugino masses lead to nonzero imaginary part of squark couplings.
On the other hand, for the non-zero imaginary part of 
the Passarino-Veltman function,
a pair of internal lines of the loop should be 
on-shell, and thus the mass of the charged Higgs should exceed 
the sum of the stop and sbottom masses, 
$m_{H^\pm}> m_{\tilde{t}}+m_{\tilde{b}}$.
Then the gluino loop and $\tilde{t}-\tilde{b}$ self-energy
contribution to ${\rm Re}~\delta~y_i^{\rm CP}$ are switched on 
and we obtain a sizable CP violating asymmetry.
It is also possible to obtain non-zero Im($C_I,~B_0$) 
when $\sqrt{\hat t}$ exceeds the sum of the gluino and stop masses,
$\sqrt{\hat t} > m_{\tilde{g}}+m_{\tilde{t}}$,
but this effect is convoluted by the parton distributions
and brings on fluctuations as shown in the figures.

The hadronic cross section is given by
\be
\sigma(p p \to g \bar{b} \to \bar{t} H^+) 
       = \int dx~dy~f_g(x)~f_b(y)~\hat \sigma(g \bar{b} \to \bar{t} H^+),
\ee
where $f_{g}$ and $f_b$ are the parton distribution functions (PDF)
for gluon and $b$-quark respectively.
We use the leading order MRST functions for a gluon and a $b$-quark 
PDF in a proton \cite{MRST}. 
The QCD factorization and renormalization scales $Q$ are set to be the
$g b$ invariant mass,  {\it i.e.}, $\sqrt{\hat{s}}$.
The $Q^2$-dependence will be small by the cancellation 
in the CP asymmetry which is of our main interest.
The center-of-momentum (c.m.) energy of $pp$ collisions at the LHC
is $\sqrt{s}=14$ TeV
and the kinematic cuts of $p_{_T} \geq 25$ GeV 
and $|\eta| \leq 2.5$ have been employed in this paper.
The CP violating asymmetry is defined at the hadronic level by
\be
A_{CP} &=& \frac{\sigma(p p \to g \bar{b} \to \bar{t} H^+)
              -\sigma(p p \to g b \to t H^-)}
              {\sigma(p p \to g \bar{b} \to \bar{t} H^+)
              +\sigma(p p \to g b \to t H^-)}.
\ee


We do not assume the specific SUSY model in our analysis 
but just fix a few parameters in order to avoid the confusion 
caused by too many parameters.
The values of parameters are taken to be
\be
&&M_2 = 200~{\rm GeV}, ~~~~ m_{\tilde{Q}} = 350~{\rm GeV},
\nonumber \\
&&|A_t| = |A_b| = 500~{\rm GeV},~~~~\mu = 500~{\rm GeV},
\ee
and the GUT relation is assumed for the gaugino masses.
The phase of $\mu$ is strictly constrained
by the electric dipole moments (EDM) of the electron and neutron
in the mSUGRA.
Although the constraint on the phase of $\mu$ 
would not be serious in other SUSY models,
we ignore the $\mu$ phase for simplicity in this work.
Consequently varied parameters are 
the charged Higgs mass, $\tan \beta$,
and phases of $A_t$, $A_b$, and $M_3$. 
For instance, our choice of parameters yields the sparticle spectrum
as follows:
\be
m_{\tilde t} &=& 208.84,~~ 473.62~~{\rm GeV},
\nonumber \\
m_{\tilde b} &=& 351.66,~~ 371.13~~{\rm GeV},
\nonumber \\
m_{\tilde \chi^0} &=& 97.70,~~ 189.04,~~ -503.81,~~ 517.27~~{\rm GeV},
\nonumber \\
m_{\tilde \chi^\pm} &=& 204.51,~~ 715.25~~{\rm GeV},
\nonumber \\
m_{\tilde g} &=& 703.65~~{\rm GeV},
\ee
for $m_{H^\pm} = 800$ GeV, $\tan \beta = 5$.
Our choice is similar to the sparticle spectra of 
the typical or the light stop scenario 
in the mSUGRA model \cite{snowmass}.

In the Fig. 2 and 3, we show the behaviors of the CP violating asymmetry
with respect to $m_{H^\pm}$ when the phases of $A_t$ and $M_3$
are nonzero.
We take $\theta_t = \pi/2$ and other phases are set to be zero
in Fig 2 and take $\theta_t = \pi$ and $\phi_3=\pi/2$ in Fig. 3
to maximize $A_{CP}$.
Since the CP violating correction ${\rm Re}~\delta y_t^{\rm CP}$
is generically larger than ${\rm Re}~\delta y_b^{\rm CP}$,
the CP violating asymmetry becomes larger as $\tan \beta$ decreases
in the plots.
We can see that the sizable $A_{CP}$ arises around 600 GeV
when $m_{H^\pm}$ exceeds $m_{\tilde{t}_1}+m_{\tilde{b}_1}$
and the Passarino-Veltman function develops
the non-zero imaginary part.
The next rising-up of the plot arises caused by the loops
with the internal lines of heavier stop
around $m_{H^\pm}=850$ GeV, that is, 
when $m_{H^\pm} > m_{\tilde{t}_2}+m_{\tilde{b}_1}$.

At a collider, what we observe is not the charged Higgs boson itself
but its decay products.
The charged Higgs boson dominantly decays into $t b$ and $\tau \nu_\tau$
pair if $m_{H^\pm} > m_t$.
The CP violating SUSY corrections to the Yukawa coupling $tbH^\pm$
also give rise to a difference between decay rates of 
$H^+ \to t \bar b$ and $H^- \to \bar t b$ 
indicating a CP violation as considered in Ref. \cite{christova}.
Therefore the measured asymmetry is actually the combined asymmetry
\be
A^{\rm tot}_{CP} &=& \frac{\sigma(p p \to g \bar{b} \to \bar{t} H^+)
                 \Gamma(H^+ \to t \bar b)
          -\sigma(p p \to g b \to t H^-) \Gamma(H^- \to \bar t b) }
          {\sigma(p p \to g \bar{b} \to \bar{t} H^+)
                 \Gamma(H^+ \to t \bar b)
          +\sigma(p p \to g b \to t H^-) \Gamma(H^- \to \bar t b) }.
\ee
Under the assumption of Re $\delta y^{\rm CP}_i \ll y_i^0$,
the asymmetry is approximately the sum of the production asymmetry
and the decay asymmetry, 
\be
A^{\rm tot}_{CP} \sim \frac{\Delta \sigma}{\sigma} 
                         + \frac{\Delta \Gamma}{\Gamma} 
                  = A^{\rm prod}_{CP} + A^{\rm decay}_{CP}.
\ee
Thus it is expected that the total asymmetry is amplified 
due to the asymmetry in decay.
We show the total asymmetry with respect to $m_{H^\pm}$ in Fig. 4.
If we consider the events with $H^\pm$ decays into $\tau \nu_\tau$,
it is enough to consider the asymmetry of Eq. (6)
since the rate asymmetry of  $H^\pm \to \tau \nu_\tau$
is nothing but of order $10^{-3}$ \cite{christovatau}.


It is known that the SUSY corrections play an important role 
in the charged Higgs production processes.
They can enhance the signal-background ratio to shift 
the reach of the LHC search by a few hundred GeV
and may yield CP violation in the charged Higgs production
\cite{belyaev}.
The QCD corrections and SUSY corrections to the gluonic vertex 
and propagators are canceled out in the CP asymmetry
and we do not consider them here.
We find that the CP violating asymmetry can reach more than 10 \%
when the mass of the charged Higgs boson is larger than 850 GeV.
If we consider the CP asymmetry of decay products,
$pp \to t \bar{t} b / \bar{t} t \bar{b}$,
it is amplified by the CP violation of decay process and
we could obtain the asymmetry more than 25 \%
as shown in Fig. 4.
In conclusion, the CP violation in the charged Higgs boson production 
associated with a top quark can be large dependent on SUSY phases 
and $m_{H^\pm}$.
Thus we expect that
the CP violating asymmetry could be detected 
in the charged Higgs boson production at the LHC.

\nopagebreak

\acknowledgments
We thank Dr. S. Baek for helpful discussions and Dr. J.S. Lee
for his valuable comments.
This work was supported by Korea Research Foundation Grant
(KRF-2003-050-C00003).




\def\PRD #1 #2 #3 {Phys. Rev. D {\bf#1},\ #2 (#3)}
\def\PRL #1 #2 #3 {Phys. Rev. Lett. {\bf#1},\ #2 (#3)}
\def\PLB #1 #2 #3 {Phys. Lett. B {\bf#1},\ #2 (#3)}
\def\NPB #1 #2 #3 {Nucl. Phys. {\bf B#1},\ #2 (#3)}
\def\ZPC #1 #2 #3 {Z. Phys. C {\bf#1},\ #2 (#3)}
\def\EPJ #1 #2 #3 {Euro. Phys. J. C {\bf#1},\ #2 (#3)}
\def\JHEP #1 #2 #3 {JHEP {\bf#1},\ #2 (#3)}
\def\IJMP #1 #2 #3 {Int. J. Mod. Phys. A {\bf#1},\ #2 (#3)}
\def\MPL #1 #2 #3 {Mod. Phys. Lett. A {\bf#1},\ #2 (#3)}
\def\PTP #1 #2 #3 {Prog. Theor. Phys. {\bf#1},\ #2 (#3)}
\def\PR #1 #2 #3 {Phys. Rep. {\bf#1},\ #2 (#3)}
\def\RMP #1 #2 #3 {Rev. Mod. Phys. {\bf#1},\ #2 (#3)}
\def\PRold #1 #2 #3 {Phys. Rev. {\bf#1},\ #2 (#3)}
\def\IBID #1 #2 #3 {{\it ibid.} {\bf#1},\ #2 (#3)}


\begin{center}
\begin{figure}[b]
\hbox to\textwidth{\hss\epsfig{file=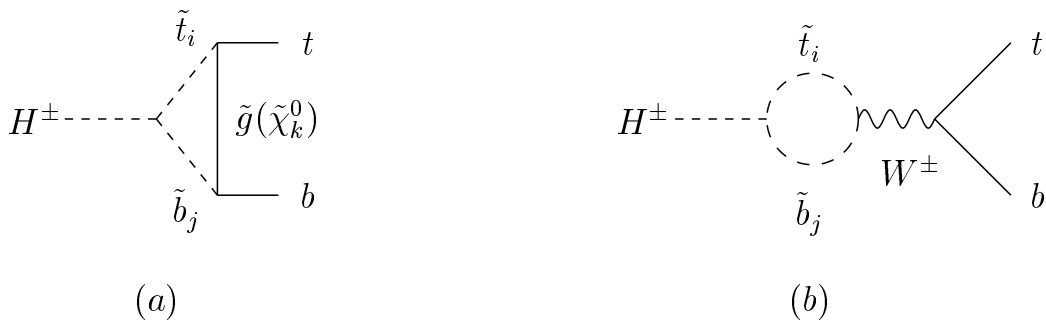,width=14cm}\hss}
\vspace{1cm}
\caption{
Diagrams of dominant SUSY correction to the $tbH^\pm$ vertex.
}
\end{figure}
\end{center}

\newpage

\begin{center}
\begin{figure}[b]
\hbox to\textwidth{\hss\epsfig{file=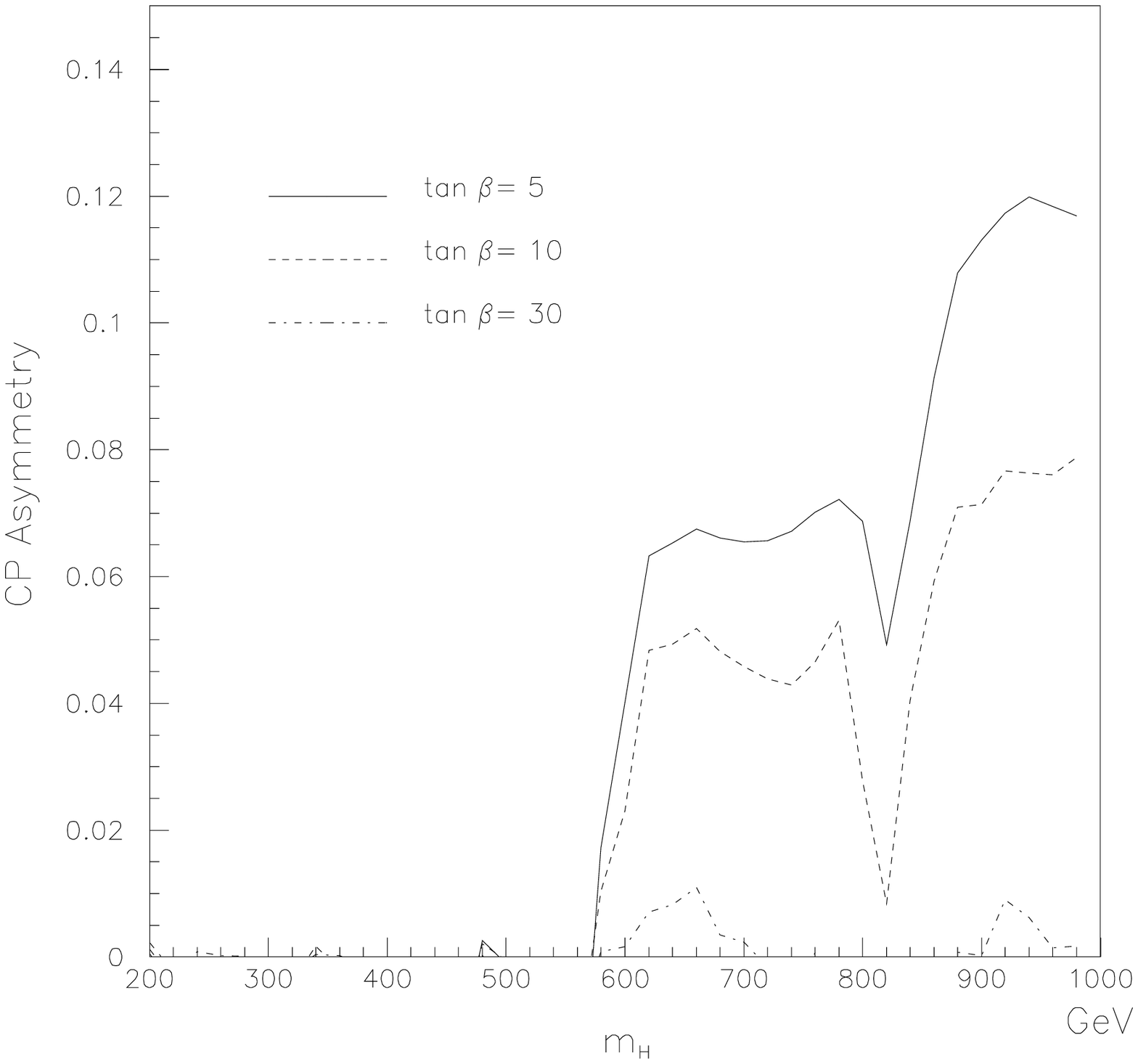,width=14cm}\hss}
\vspace{1cm}
\caption{
CP violating asymmetry due to the SUSY correction
with varying $m_{H^\pm}$
and $\theta_t = \pi/2$.
Other phases are set to be 0.
The solid line denotes $\tan \beta = 5$,
the dashed line $\tan \beta = 10$, 
and the dash-dotted line $\tan \beta = 30$.
}
\end{figure}
\end{center}

\newpage

\begin{center}
\begin{figure}[b]
\hbox to\textwidth{\hss\epsfig{file=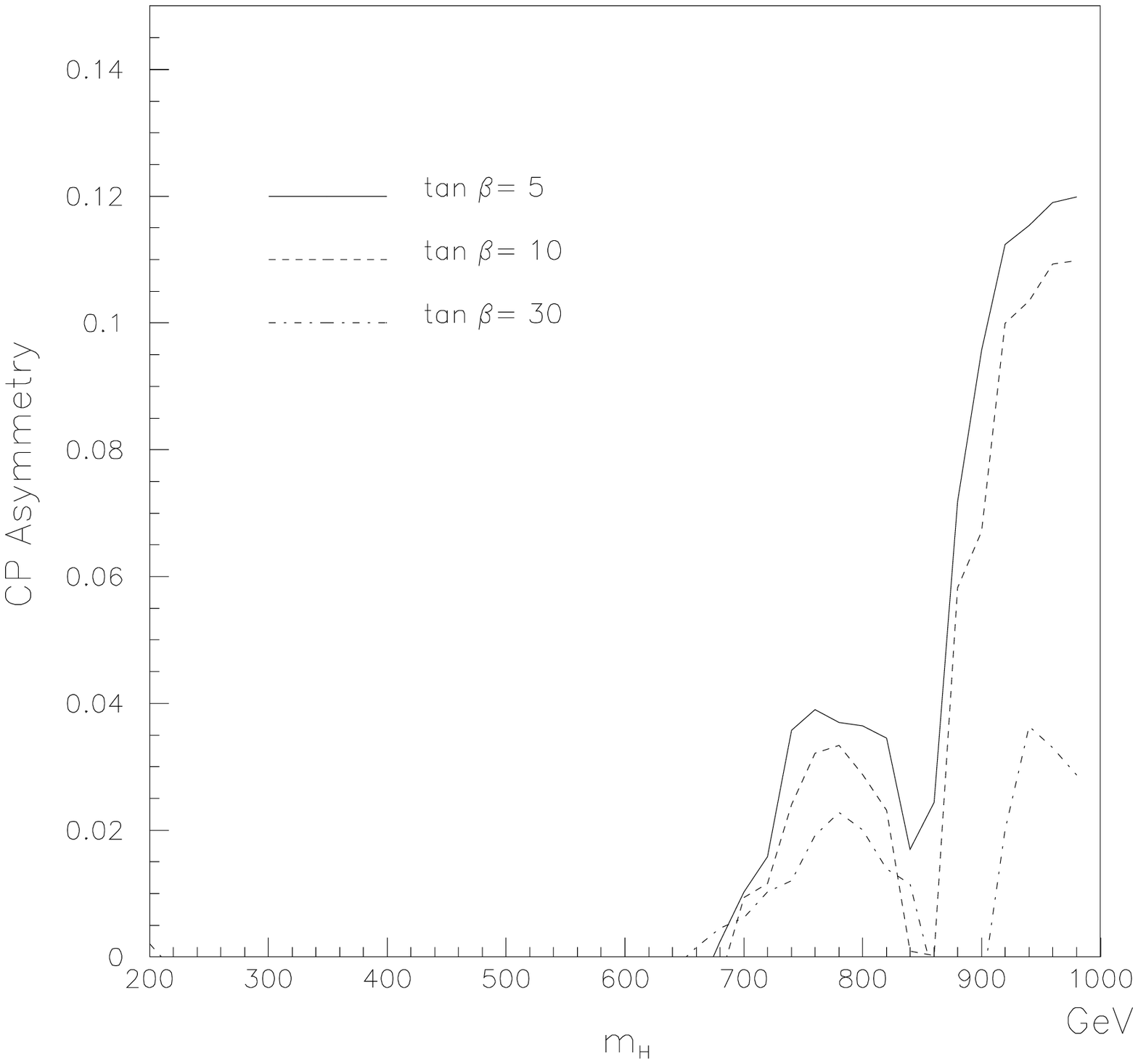,width=14cm}\hss}
\vspace{1cm}
\caption{
CP violating asymmetry due to the SUSY correction
with varying $m_{H^\pm}$
and  $\theta_t = \pi$, $\phi_3 = \pi/2$.
Other phases are set to be 0.
The solid line denotes $\tan \beta = 5$,
the dashed line $\tan \beta = 10$, 
and the dash-dotted line $\tan \beta = 30$.
}
\end{figure}
\end{center}

\newpage

\begin{center}
\begin{figure}[b]
\hbox to\textwidth{\hss\epsfig{file=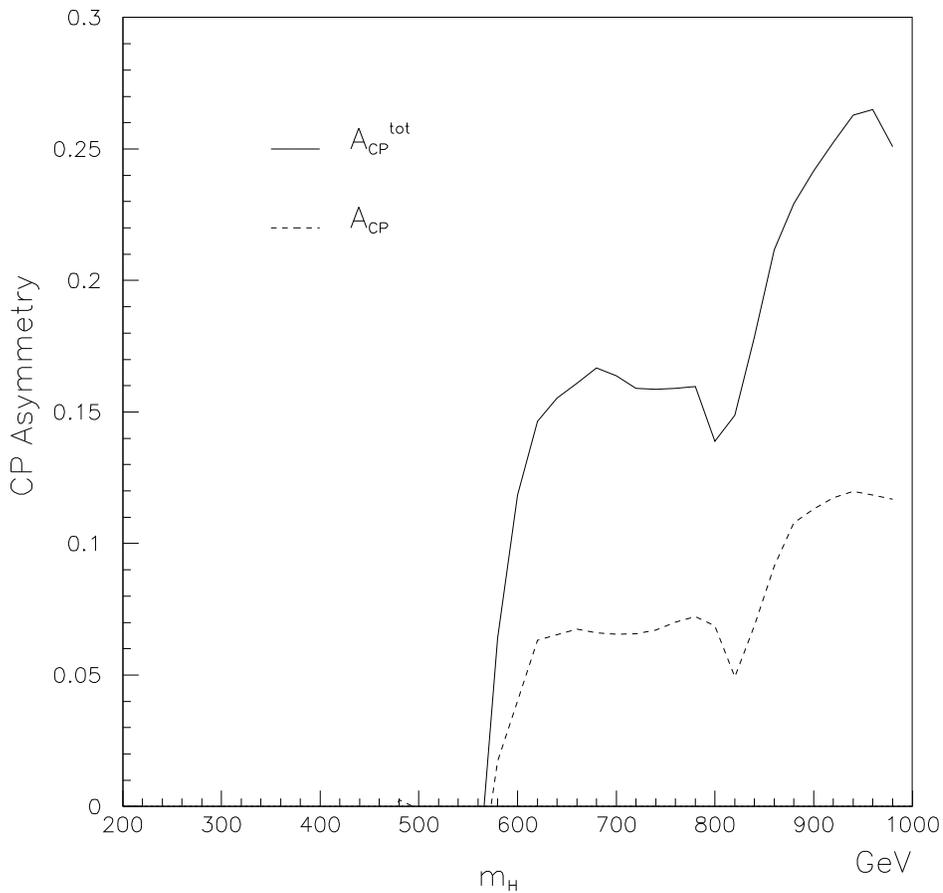,width=14cm}\hss}
\vspace{1cm}
\caption{
Total CP violating asymmetry and production CP asymmetry
due to the SUSY correction with varying $m_{H^\pm}$
and $\theta_t = \pi/2$.
The solid line denotes the total asymmetry for the $H^\pm$ production
and the decay into $tb$ pair.
The dashed line denotes the asymmetry for the $H^\pm$ production only.
}
\end{figure}
\end{center}

\end{document}